\documentclass[twocolumn, showpacs,%
  nofootinbib,aps,superscriptaddress,%
  prd,notitlepage,showkeys,10pt]{revtex4-1}

\usepackage{amssymb}
\usepackage{amsmath}
\usepackage{graphicx}
\usepackage{dcolumn}
\usepackage{hyperref}

\begin{document}

\title{Comment on ``Particle Path Through a Nested Mach-Zehnder Interferometer''}
\author{Hatim Salih}
\email[]{hatim.salih@york.ac.uk or salih.hatim@gmail.com}
\affiliation{Department of Mathematics, University of York, Heslington, York YO10 5DD, UK}

\date{\today}

\begin{abstract}
In a recent paper, Phys. Rev. A 94, 032115 (2016), Griffiths questioned---based on an interesting consistent-histories (CH) argument---the counterfactuality, for one of the bit choices, of Salih et al.'s protocol for communicating without sending physical particles, Phys. Rev. Lett. 110, 170502 (2013). Here, we first show that for the Mach-Zehnder version used to explain our protocol, contrary to Griffiths's claim, no family of consistent histories exists where any history has the photon traveling through the communication channel, thus rendering the question of whether the photon was in the communication channel meaningless from a CH viewpoint. We then show that for the actual Michelson-type protocol, there is a consistent-histories family for each cycle that includes histories where the photon travels through the communication channel. We show that the probability of finding the photon in the communication channel at any time is zero---proving complete counterfactuality.

\end{abstract}

%\pacs{03.65.Ta, 42.50.Xa}

\maketitle

The protocol in question for direct counterfactual quantum communication is explained in \cite{Salih} using nested Mach-Zehnder interferometrs at first, much easier to explain, before the actual protocol is given in Michelson form. The Michelson implementation captures the key functionality of its Mach-Zehnder counterpart while allowing a substantial saving of physical resources. But, as we explain below, not only is the Michelson implementation more powerful in this sense, it also allows couterfactuality to be verified in a way that may not be possible using Mach-Zehnder interferometry.

The debate about counterfactuality for the bit value corresponding to Bob not blocking the communication channel can be explained using only two nested Mach-Zehnder interferometers. (Counterfactuality is not questioned for the bit value corresponding to Bob blocking the channel.) We reproduce the key figure from \cite{Griffiths} in our FIG. \ref{Figure1} below. In order to frame the debate in the context of counterfactual communication, imagine the two nested interferometers of FIG. \ref{Figure1} rotated by 45 degrees clockwise with the interferometers and detectors thought of as being in Alice's station on the left---except arm C, which would correspond to the communication channel leading to Bob on the right.

By constructing a family of consistent histories that includes a history where the photon takes path $A$ between $S$ at time $t_0$, and $F$ at time $t_4$, we can ask what the probability of the photon taking path $A$ is. As shown in \cite{Griffiths}, such a family exists, namely, $\mathcal{F}'_A: S_0 \odot \left \{A_1,D_1,Q_1  \right \} \odot \left \{A_2,B_2+C_2  \right \} \odot \left \{A_3,E_3,H_3  \right \} \odot F_4$, where $S_0$ is the projector onto arm $S$ at time $t_0$, $A_1$ is the projector onto arm $A$ at time $t_1$, etc.. Using the extended Born rule, the probability of the photon taking path $A$ is one, meaning that the photon remains in Alice's domain at all times. What about the probability of finding the photon in the communication channel, $C$? In general, as shown in \cite{Griffiths}, there is no consistent histories family that includes a history where the photon travels through $C$, rendering the question meaningless---except for a special choice of reflectivity for mirrors 1 and 4. This reflectivity, equal to $1/3$, however, lies outside the parameter space of Salih et al.'s protocol which has reflectivity $\cos^2 \frac{\pi}{2M}$ for beamsplitters of outer interferometer, and reflectivity $\cos^2 \frac{\pi}{2N}$ for beamsplitters of inner interferometer, with $M$ and $N$ both ranging from 2 for the smallest number of interferometers, giving reflectivity $1/2$, to very large, giving reflectivity asymptotically close to one in the limit. In fact a reflectivity of $1/3$ would correspond to a probability of correctly guessing Bob's bit choice when Bob does not block the channel equal to $1/3$, which is worse than random guessing. In other words there is no consistent histories family for Salih et al.'s Mach-Zehnder protocol with the photon traveling through the communication channel.

However, this is not the only reason why such a family does not exist. The basis of the counterfactuality of Salih et al.'s protocol, as stated in \cite{Salih reply}, is that ``the probability of the photon existing at location E is zero'', which the analysis in section 7 in \cite{Griffiths} does not take account of. In particular, Griffiths's proposition $P1$, ``The particle was in $S$ at $t_0$ and in $F$ at $t_4$'' does not include ``the particle was not in E at $t_3$''. $\mathcal{F}_C$ has to be refined to include events at time $t_3$, which, as shown in \cite{Griffiths}, will make it inconsistent regardless of mirror reflectivity choice. Whereas one can say based on CH that the photon remains in channel $A$ in Alice's domain at all times, one cannot say that the photon was (or wasn't) in the communication channel at any time.

We now give for the actual Michelson protocol a consistent-histories family for each outer cycle {\it separately}, that includes histories where the photon travels through the communication channel, before we show that the probability of finding the photon in the communication channel at any point is zero. FIG. \ref{EquMichelson} is equivalent to one outer cycle of the Michelson version of the protocol \cite{Salih}, which includes two inner cycles, laid out sequentially in time for clarity, for the case in question of Bob not blocking the channel, with $M=N=2$, where $M$ and $N$ are the number of inner and outer cycles respectively. Note that an outer cycle starts in channel $S$ at the top of FIG. \ref{EquMichelson}, branching into $A$ and $D$, before ending in $S$ again at the bottom. Here, each outer cycle has two nested inner cycles within it, where an inner cycle starts in channel $D$, branching into $B$ and $C$, before ending in $D$ again. Consider the first outer cycle. As can be seen from FIG. \ref{EquMichelson}, the photon starts, at time $t_0$, in channel $S$, $H$-polarised. And given that it is not lost to the detector at the bottom, it emerges at time $t_4$ in channel $S$, always $H$-polarised. (The basic idea behind the protocol is that if Bob instead blocks the channel then the photon emerging in $S$ would have a $V$-polarised component that is amplified by successive outer cycles.) The same applies to the second outer cycle, and more generally to any outer cycle for $M>2$ and $N >2$.

Let us start with the following consistent histories family for either of the two outer cycles. Note that here we are looking at one outer cycle only, with $t_1$ corresponding to the begging of the outer cycle (which is not necessarily the begging of the protocol) and $t_4$ corresponding to the end on the outer cycle (which is not necessarily the end of the protocol).

\begin{align}
& \mathcal{Y}: S_0 \otimes I_0 \odot I_1 \otimes I_1 \odot I_2 \otimes I_2 \odot I_3 \otimes I_3 \odot S_4 \otimes H_4,\nonumber \\
& \mathcal{Y}': S_0 \otimes I_0 \odot I_1 \otimes I_1 \odot I_2 \otimes I_2 \odot I_3 \otimes I_3 \odot S_4 \otimes V_4,\nonumber \\
& \mathcal{Y}'': \widetilde{S}_0 \otimes I_0 \odot I_1 \otimes I_1 \odot I_2 \otimes I_2 \odot I_3 \otimes I_3 \odot I_4 \otimes I_4, \nonumber \\
& \mathcal{Y}''': S_0 \otimes I_0 \odot I_1 \otimes I_1 \odot I_2 \otimes I_2 \odot I_3 \otimes I_3 \odot \widetilde{S}_4 \otimes I_4 \nonumber
\end{align}

where $S_0$ and $I_0$ are channel and the polarisation projectors respectively at time $t_0$, $I$ being the identity projector, and $\widetilde{S}$ the projector onto the complement of channel S, and so on for other times. Because we already know that the photon always starts at time $t_0$ in channel $S$, we can omit from the discussion history $\mathcal{Y}''$ whose channel projector at $t_0$ is $\widetilde{S}$. And because we assume the photon is not lost (to detector), i.e. it is in channel $S$ at time $t_4$, we can also omit from the discussion history $\mathcal{Y}'''$ whose channel projector at $t_4$ is $\widetilde{S}$.

Also, since the identity polarisation projector $I_0=H_0+V_0$, and because we already know that the photon is $H$-polarised at $t_0$, we can rewrite histories $\mathcal{Y}$ and $\mathcal{Y}'$ as,

\begin{align}
& \mathcal{Y}: S_0 \otimes H_0 \odot I_1 \otimes I_1 \odot I_2 \otimes I_2 \odot I_3 \otimes I_3 \odot S_4 \otimes H_4, \nonumber \\
& \mathcal{Y}': S_0 \otimes H_0 \odot I_1 \otimes I_1 \odot I_2 \otimes I_2 \odot I_3 \otimes I_3 \odot S_4 \otimes V_4 \nonumber
\end{align}

But the chain ket associated with history $\mathcal{Y}'$, namely $\left| S_0 \otimes H_0, I_1 \otimes I_1, I_2 \otimes I_2, I_3 \otimes I_3, S_4 \otimes V_4 \right\rangle$, whose inner product with itself gives by the extended Born rule the probability of finding the photon at time $t_4$ in $S$, $V$-polarised, is clearly zero as we already know from FIG. \ref{EquMichelson} that if the photon is found in $S$ at time $t_4$ it is always $H$-polarised. This can be directly verified as $\left| S_0 \otimes H_0, I_1 \otimes I_1, I_2 \otimes I_2, I_3 \otimes I_3, S_4 \otimes V_4 \right\rangle = (S_4 \otimes V_4) T_{4,3} (I_3 \otimes I_3) T_{3,2} (I_2 \otimes I_2) T_{2,1} (I_1 \otimes I_1) T_{1,0} (S_0 \otimes H_0)=0$, where $T_{1,0}$ is the corresponding unitary transformation between times $t_0$ and $t_1$, etc. \cite{Griffiths}. This only leaves history $\mathcal{Y}$.

We now refine history $\mathcal{Y}$ independently of other histories without loss of consistency,

\begin{align}
\mathcal{Y}: & S_0 \otimes H_0 \odot \left \{ A_1 \otimes I_1, D_1 \otimes I_1 \right \} \odot \nonumber \\ & \left \{ A_2 \otimes I_2, B_2 \otimes I_2, C_2 \otimes I_2, \right \} \odot \nonumber \\ & \left \{ A_3 \otimes I_3, B_3 \otimes I_3, C_3 \otimes I_3, \right \} \odot S_4 \otimes H_4 \nonumber
\end{align}

This histories family is consistent as each of its 18 chain kets is {\it zero}, except the chain ket $\left| S_0 \otimes H_0, A_1 \otimes I_1, A_2 \otimes I_2, A_3 \otimes I_3, S_4 \otimes H_4 \right\rangle$, and are therefore mutually orthogonal. For example, the chain ket $\left| S_0 \otimes H_0, D_1 \otimes I_1, C_2 \otimes I_2, C_3 \otimes I_3, S_4 \otimes H_4 \right\rangle = (S_4 \otimes H_4) T_{4,3} (C_3 \otimes I_3) T_{3,2} (C_2 \otimes I_2) T_{2,1} (D_1 \otimes I_1) T_{1,0} (S_0 \otimes H_0) = \frac{1}{\sqrt{2}} (S_4 \otimes H_4) T_{4,3} (C_3 \otimes I_3) T_{3,2} (C_2 \otimes I_2) T_{2,1} (D_1 \otimes V_1) = \frac{1}{2} (S_4 \otimes H_4) T_{4,3} (C_3 \otimes I_3) T_{3,2} (C_2 \otimes H_2)= \frac{1}{2\sqrt{2}} (S_4 \otimes H_4) T_{4,3} (C_3 \otimes H_3) = \frac{1}{2\sqrt{2}}(S_4 \otimes H_4) (\widetilde{S}_4 \otimes H_4)=0$. We conclude based on consistent histories that during either of the two identical cycles, the first starting at $t_0$ and ending at $t_4$, and the second starting at the equivalent of $t_0$ and ending at the equivalent of $t_4$, the photon was not in the communication channel, and was therefore not in the communication channel at any time. This straightforwardly extends to $M>2$ and $N >2$.

Note that the above CH analysis does not violate the single framework rule, which tells us that incompatible frameworks give rise to conceptual problems when one tries to apply them to the same system during the same time interval, Section 16.4 of Ref. \cite{Griffiths book}. Our analysis applies two (equivalent) frameworks, i.e. two (equivalent) CH families, at {\it different} time intervals. To see that this does not violate the single framework rule, consider the two approaches given in the preceding paragraph in Section 16.4 of Ref. \cite{Griffiths book} for combining conclusions drawn based on two, even incompatible frameworks, ``The conceptual difficulty goes away if one supposes that the two incompatible frameworks are being used to describe two distinct physical systems that are described by the same initial data, {\it or the same system during two different runs of an experiment}'' (Italics mine). Since in the protocol we are analysing, each outer cycle is identical, we are effectively looking at the same system during different runs of the experiment. Conclusions can therefore be combined without conceptual problems. The photon was not in the communication channel during the first outer cycle, it was not in the communication channel during the second outer cycle, therefore it was not in the communication channel at any time.

Interestingly, if one considers histories for the two-outer-cycles two-inner-cycles case, with the initial state taken to be at the start of the first outer cycle, and the final state taken to be at the end of the protocol, after the second outer cycle, then we have the following paradoxical situation. If one refines the histories family to enable asking whether the photon was in the communication channel during the second outer cycle only, then by a similar argument to the above, the histories family is consistent, giving zero probability for the photon having been in the communication channel. On the other hand, if one refines the histories family to enable asking whether the photon was in the communication channel during the first outer cycle, then (because of an artifact arising from the action of $PR_1$ in the second outer cycle) the histories family is not consistent, rendering it meaningless to ask about the probability of the photon having been in the communication channel! But these two cycles are identical: the photon starts each cycle in $S$, $H$-polarised and ends each cycle in $S$, $H$-polarised, having undergone the exact same transformations. Our above approach of constructing a consistent-histories family per cycle resolves this paradox. (A detailed explanation of the paradox is left to a future work.)

In summary, we have constructed a consistent-histories family for each of the identical cycles of the actual Michelson protocol of Salih et al., separately, for the case in question of Bob not blocking the channel, allowing one to ask whether the photon was in the communication channel at any point. Since the probability is provably zero for the photon being in the communication channel between the time corresponding to the beginning of each cycle and the time corresponding to the end of the cycle, we have shown that the photon was not in the communication channel at any time, proving complete counterfactuality. 

\begin{acknowledgments}
I thank Robert Griffiths for useful comments, and for suggesting the sequential-layout-in-time for FIG. \ref{EquMichelson}. This work has been supported by Qubet Research, a start-up in quantum communication technology.

\end{acknowledgments}

%\section*{Appendix} 

\clearpage

\begin{figure}
\centering
\includegraphics[width=0.5\textwidth]{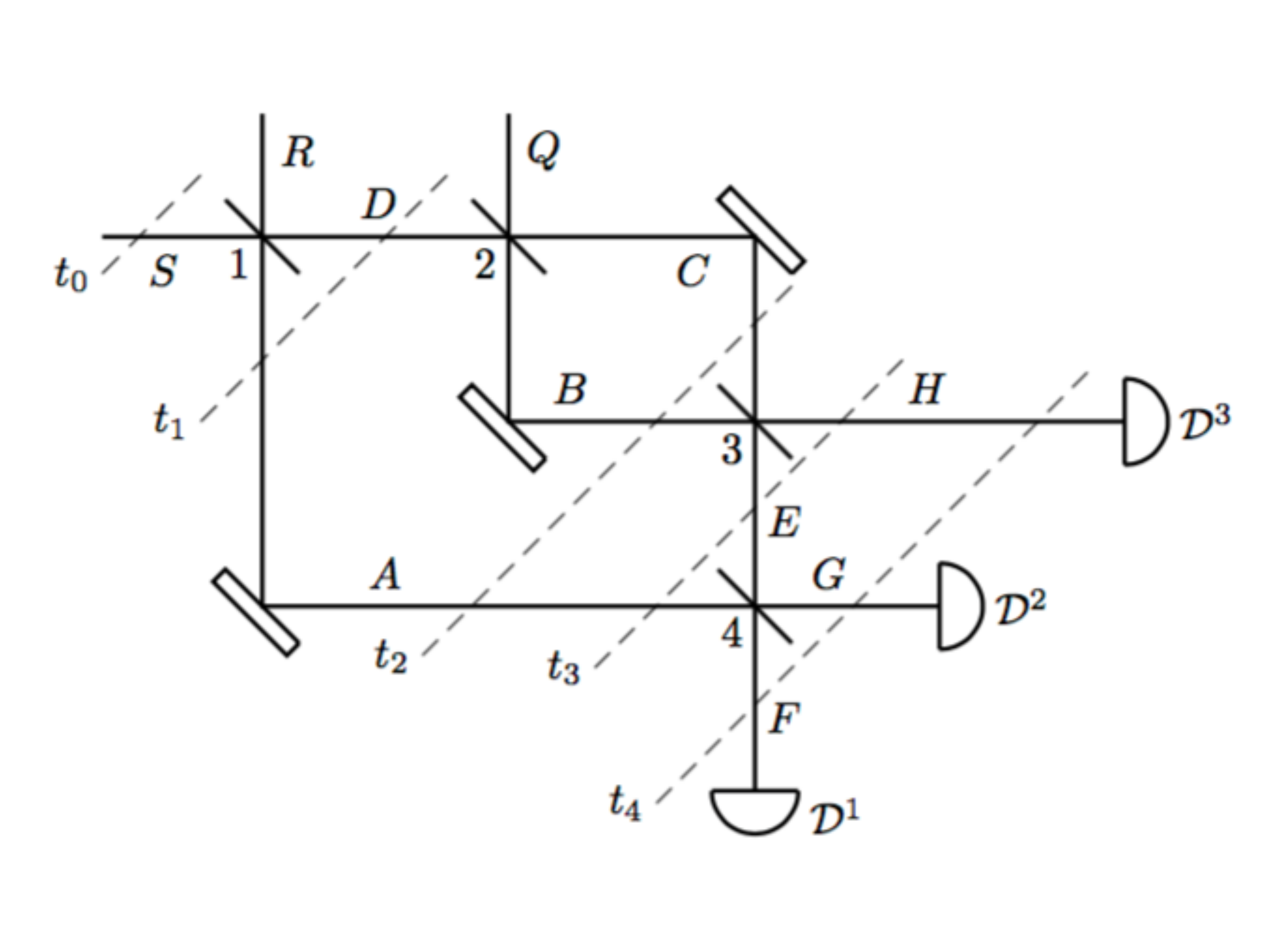}
\caption{\label{Figure1}This diagram, reproduced from \cite{Griffiths}, captures the essence of the Mach-Zehnder counterfactual communication protocol of Salih et al. \cite{Salih}. One has to picture the two nested interferometers rotated by 45 degrees clockwise, with the whole setup thought of as belonging to Alice with the exception of path $C$, which would correspond to the communication channel, and the mirror therein (double lines), which would belong to Bob. We adopt Griffiths's definitions from reference \cite{Griffiths}: ``The tilted solid lines are beam splitters numbered 1, 2, 3, 4; the double tilted lines are mirrors; the semicircles are detectors. The horizontal and vertical lines indicate different channels which are possible particle (photon) paths. The reflectivities and phases of beam splitters 2 and 3 associated with the inner MZI are chosen so that a particle entering through D will exit through H and be detected by $D^3$, rather than passing into E. The intersections of the dashed lines with the particle paths indicate possible locations of the particle at the successive times $t_0 < t_1 < t_2 < t_3 < t_4$''.}
\end{figure}

\begin{figure}
\centering
\includegraphics[width=1.0\textwidth]{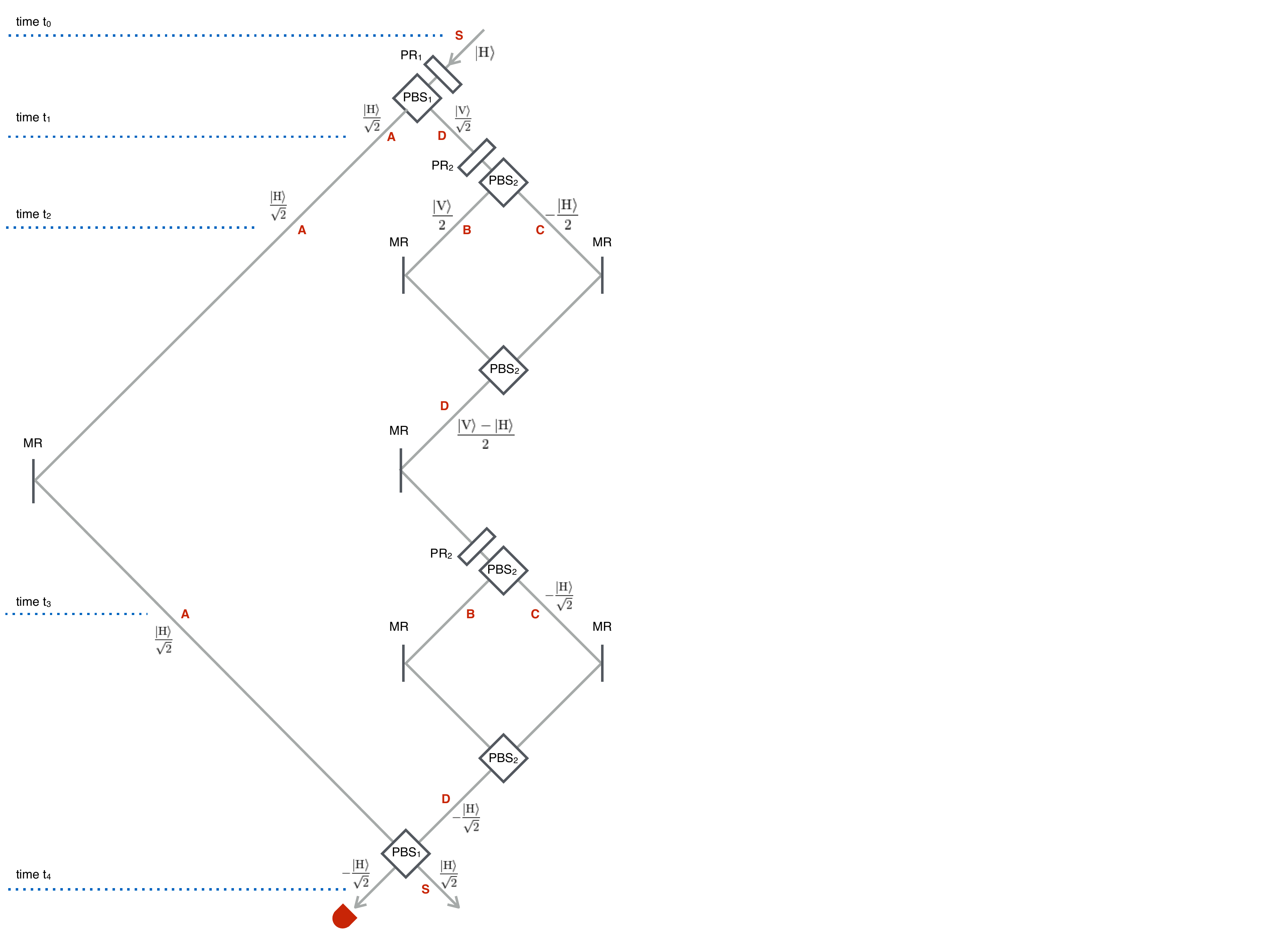}
\caption{\label{EquMichelson}The equivalent of the action of one outer cycle, and two inner cycles, of the Michelson chained quantum Zeno effect (CQZE) \cite{Salih}, laid out sequentially in time for clarity, for the case of Bob not blocking the channel, with $M=N=2$, where $M$ and $N$ are the number of inner and outer cycles respectively. The photon always starts the outer cycle in channel $S$ at the top, $H$-polarised, and finishes the cycle in channel $S$ at the bottom, $H$-polarised, provided it is not lost to the detector at the bottom. Here, $PBS$ stands for polarising beam-splitter, $PR$ stands for polarisation rotator, and $MR$ stands for mirror. Note that whereas $PR$ was applied twice per cycle in \cite{Salih} for practicality (to avoid having to switch it on and off), it is applied here once per cycle for clarity, as in \cite{Salih2014a, Salih2014b}.}
\end{figure}

\end{document}